# ARPES experiment in fermiology of quasi-2D metals (Review Article)

## A. A. Kordyuk

*G.V. Kurdyumov Institute of Metal Physics of the National Academy of Sciences of Ukraine,*
*36 Vernadsky St., Kiev 03142 Ukraine*

Angle resolved photoemission spectroscopy (ARPES) enables direct observation of the Fermi surface and underlying electronic structure of crystals—the basic concepts to describe all the electronic properties of solids and to understand the key electronic interactions involved. The method is the most effective to study quasi-2D metals, to which the subjects of almost all hot problems in modern condensed matter physics have happened to belong. This has forced incredibly the development of the ARPES method which we face now. The aim of this paper is to introduce to the reader the state-of-the-art ARPES, reviewing the results of its application to such topical problems as high temperature superconductivity in cuprates and iron based superconductors, and electronic ordering in the transition metal dichalcogenides and manganites.

Contents

## 1. Introduction

In the last century, the electronic band structure and, in particular, the Fermi surface were extremely valuable, but purely theoretical concepts introduced in order to explain many physical properties of solids, such as electrical conductivity, magnetoresistance, and magnetization oscillations upon varying magnetic field [1]. Nowadays, due to the development of new powerful experimental techniques, these concepts have become direct observables of the quantum nature of electrons in crystals. The angle resolved photoemission spectroscopy (ARPES) [2,3], which allows to see directy both the Fermi surface and the underlying electronic structure of the samples (see, e.g., Figs. 1–3), possesses a special place among these techniques.

Certainly, to "see" does not mean to "understand" and the apparent simplicity of interpretation does not guarantee validity of the conclusions. However, it can be argued that due to its visuality and vast amount of experimental data, the modern ARPES experiment provides exactly the type of everyday experience that the founders of quantum mechanics lacked so much. The purpose of this review is to show how the new opportunity to "see" in the space of electronic states, which is provided by this method, helps us to resolve the most challenging problems of solid state physics.

## 2. Modern ARPES experiment

### 2.1. ARPES spectrum

One can say that ARPES allows us to "see" the actual distribution of electrons in the energy-momentum space of the crystal. This distribution is determined by the electronic band structure (a result of interaction of the electrons with periodic crystal potential) as well as by the interaction of the electrons with inhomogeneities of the crystal lattice (defects, impurities, thermal vibrations) and with each other. In other words, in addition to the band structure, which is a consequence of the one-particle approximation, the results of the field theory methods in the many body problem, the quasiparticle spectral function and the self-energy, have become directly observable [7–9].

Photoelectron spectroscopy is based on the photoelectric effect, which has been known for more than a century [10]. Einstein has received the Nobel Prize for its explanation [11]. If the surface of a crystal is irradiated by monochromatic light of certain energy, it will emit electrons with different kinetic energy in different directions. The flux of electrons as a function of energy is called *photoemission spectrum* [12], and, when the angle of emission is resolved, it is called *ARPES spectrum*.

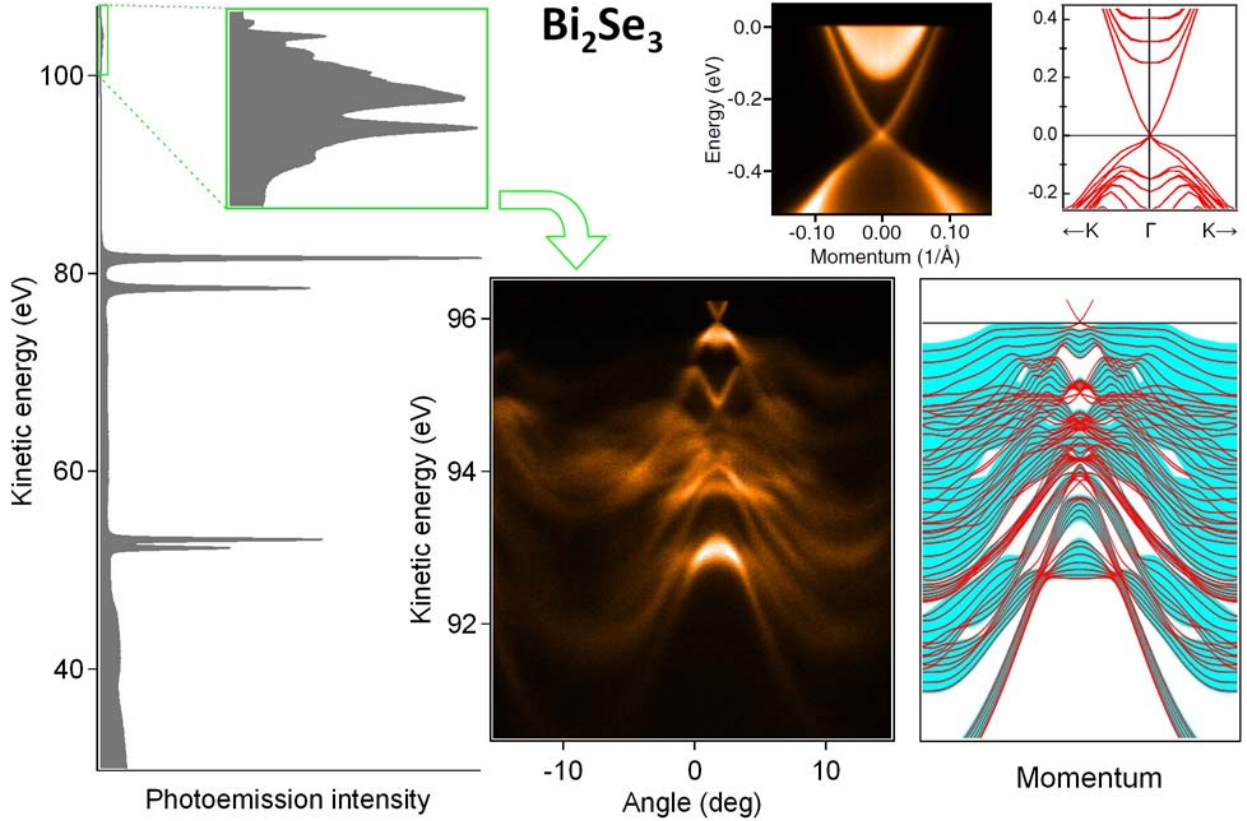

FIG. 1. ARPES spectra and electronic band structure of the topological insulator $Bi_2Se_3$ [4]: the integrated spectrum (left) recorded at $hv = 110$ eV, the valence band with angular resolution (lower row, center) recorded at $hv = 100$ eV, and the surface states in the shape of the Dirac cone [5] recorded at $hv = 20$ eV. Band calculations were performed by Krasovskii [6].

Fig. 1 (left) shows the photoemission spectrum of $Bi_2Se_3$, known as a "topological insulator". The observed peaks correspond to localized electronic levels ("core levels"), while in order to relate the electronic structure and the electronic properties of crystals, it is of utmost interest to study the states in the vicinity to the highest occupied level—the Fermi level. This region is called the valence band (enlarged in the inset above) and, when the emission angle of electrons is resolved, reveals a complex structure, which, as can be seen from a comparison with the band structure calculations, reflects the structure of the dispersive bands.

Looking ahead, it is worth mentioning that the ARPES spectrum is not just a set of one electron bands, but the single particle spectral function, which is the imaginary part of the Green's function for one-electron excitations (quasiparticles): $A(\omega, \mathbf{k}) = -\pi^{-1}\mathrm{Im}G(\omega, \mathbf{k})$ [7,8]. In the absence of interaction between the electrons, one particle states are well defined, $G_0 = 1/(\omega - \varepsilon - i0)$ and $A(\omega, \mathbf{k}) = \delta[\omega - \varepsilon(\mathbf{k})]$, where $\varepsilon(\mathbf{k})$ is the dispersion of "bare" (i.e., non-interacting) electrons. Taking into account the interaction in the normal (gapless) state, the Green's function also has a simple form $G = 1/(\omega - \varepsilon - \Sigma)$, and

$$A(\omega, \mathbf{k}) = -\frac{1}{\pi}\frac{\Sigma''(\omega)}{(\omega - \varepsilon(\mathbf{k}) - \Sigma'(\omega))^2 + \Sigma''(\omega)^2}, \quad (1)$$

where $\Sigma = \Sigma' + i\Sigma''$ is the quasiparticle self-energy, which reflects all the interactions of electrons in the crystal. Thus, not only the structure of one electron bands, but also the structure of basic interactions in the electronic system might be expected to be found in ARPES spectrum.

Comparison between experiment and calculation is a key point of this article. Since there is anyway no rigorous explanation of photoemission (commonly used "three-step" and "one-step" models [12] are, respectively, more and less rough approximations [9]), the way from experiment to model seems to be the most consistent. To paraphrase an English proverb about duck: if ARPES spectrum looks like a spectral function and behaves like a spectral function, it must be a spectral function. Moreover, a huge amount of experimental data makes such empirical approach extremely convincing. Nevertheless, to link the experimental space of angles and kinetic energy with the energy–momentum space of the crystal, explain in simple terms why this refers to spectral function, and name other components of the spectrum, let us consider the process of photoemission in more details, using some of the ideas of the mentioned above three-step model.

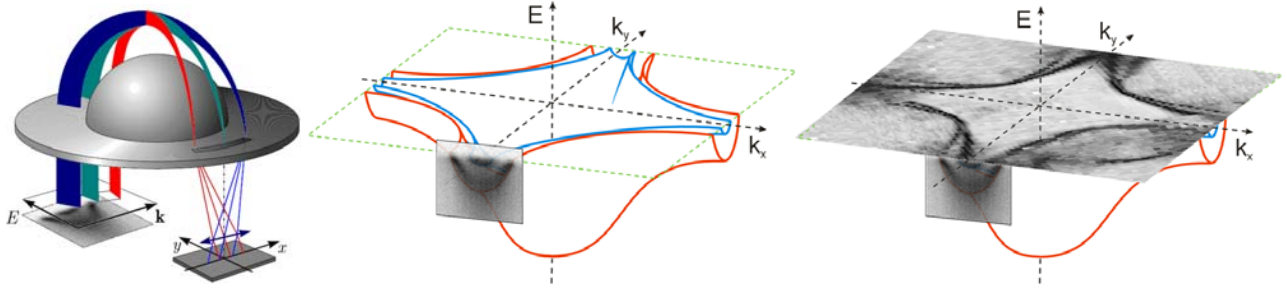

FIG. 2. The detector of a modern photoelectron analyzer (left) acts as a window into the 3D energy-momentum space of a 2D metal (center). By moving this "window" in (ω, **k**)-space, a complete distribution of electrons and, in particular, its cross-section by the Fermi level, the Fermi surface (right), can be obtained. Here, the electronic band structure and the ARPES spectra are shown for high-temperature superconductor Bi-2212.

A typical ARPES setup consists of an electron lens, a hemispherical analyzer, and a multichannel plate detector (Fig. 2, left). In the angle resolved mode, a spot of focused UV excitation on the surface of the sample (typically of the order of hundreds micrometers in diameter) coincides with the focal point of the electron lens, which projects the photoelectrons onto the entrance slit of the analyzer. Thus, the lens translates the angular space of photoelectrons into the coordinate space by forming along the slit an angular scan of electrons which fly within the plane formed by the axis of the lens and the slit. Pathing through the analyzer further, the electron beam also spreads in energy in a plane perpendicular to the slit. As a result, a two-dimensional spectrum is formed on the 2D-detector (e.g., a microchannel plate): the intensity of photoelectron emission as a function of its kinetic energy and emission angle.

To relate this spectrum to the picture of the band structure, it is assumed that an electron escaping the crystal concerves its momentum and energy. More rigorously: Upon escaping the crystal, the electron quasi-momentum $\hbar\mathbf{k}$ becomes real momentum in free space $\mathbf{p} = \hbar\mathbf{k} + \hbar\mathbf{G}$, where $\mathbf{G}$ — is any reciprocal-lattice vector. The energy conservation law takes the form: $h\nu = E_k + E_b + \phi$, i.e., the energy of an incident photon $h\nu$ is spent on overcoming the binding energy $E_b$ of an electron in the crystal and the work function $\phi$ for escaping from the sample to the analyzer, and the remainder is converted into the kinetic energy of the photoelectron $E_k = p^2/2m$.

These conservation laws can be applied to an ARPES experiments with the following reservations: (1) the momentum conservation law (quasi-momentum transformation law) applies to the component of the momentum parallel to the surface of the crystal, $\mathbf{k}_{\|}$, along which a translational symmetry takes place; (2) the component of the photon momentum along the surface is usually neglected; (3) it is assumed that the sample (with metallic conductivity) has the same potential as that of the analyzer, and therefore $\phi$ does not depend on the sample and is the work function of the analyzer. Moreover, the momentum that an electron had in the crystal is defined as $\hbar\mathbf{k}_{\|} + \hbar\mathbf{G} = \sqrt{2mE_k}\,\sin\theta$, where $\theta$ electron emission angle with respect to the normal to the surface, and the energy $\omega = -E_b = E_k + \phi - h\nu$.

From this one can assume that the ARPES-spectrum reflects the probability to find an electron in the crystal with a certain energy ω and momentum $\mathbf{k}_{\|}$ (hereafter **k**), and put it into an excited state. The first process is determined by the density of occupied states, or spectral function, multiplied by the Fermi distribution: $A(\omega, \mathbf{k})f(\omega)$. The second process is related to the probability of photon absorption, or the direct transition to the free level (in the three-step model of photoemission), which is commonly called "matrix elements" $M(h\nu, n, \mathbf{k})$. Thus, the structure of an ARPES-spectrum consisting of *n* bands can be written in the coordinates (ω, **k**) as follows:

$$\text{ARPES}(\omega, \mathbf{k}) \propto \sum_n M(h\nu, n, \mathbf{k}) A(\omega, \mathbf{k}) f(\omega). \quad (2)$$

This equation is written for the two-dimensional case, when the "bare" bands can be represented as surfaces in the 3D (ω, **k**)-space, and the Fermi surface can be represented as the contours at the Fermi level—the (0, **k**)-plane (see Figs. 2, 3, and 5). The experimental factors, such as the resolution [13] and the efficiency of the detector channels [5] are also omitted here. This is done for the sake of simplicity and also because the notion of "modern ARPES" includes a set of procedures for dealing with these factors [5,13]. Nevertheless, it is worth remembering that the underestimation of experimental factors often led to wrong conclusions, as will be shown in Section 3 for the case of cuprates.

### 2.3. Matrix elements

One should say more about the "matrix elements." In our definition, $M(h\nu, n, \mathbf{k})$ corresponds more to the probability of the one-step transition of an electron from its initial state in the crystal to the final state on its way to of the spectrometer [16,17]. For better clarity, it is separated into the probability of "photoionization" of an electron in the crystal [12] (classical photoemission matrix elements, the first step of the three-step model) and the probability of the subsequent interaction of the photoelectron with the crystal and its surface (the other two steps). However, the majority of studies still consider only the photoionization probability since it is easier to understand on which parameters it depends and to estimate this dependence [18]. In the process of photoionization, due to the interaction with an electromagnetic wave, an electron passes from an initial state, which is determined by the momentum **k** and the band number $n$, to a final state, which differs from the initial one by the energy $h\nu$. Accordingly, it is expected that the probability of photoionization depends on these parameters and the experimental geometry (namely, the polarization and orientation of the light wave with respect to the crystallographic planes of the sample). In this case the characteristic scale of changes of $M$ with respect to $h\nu$ and **k** is determined by the broadening of the final state and the dispersion of the respective bands. In the three-step model, this broadening can be deduced from the lifetime of the final state, while in the one-step model it can be found from the decay depths of the final-state wave function in the crystal [16]. Both approaches lead to the same estimate of the characteristic scale of variation of $M(h\nu, \mathbf{k})$ (the minimum peak width) of 2–5 eV in terms of energy and some fraction of the Brillouin zone in terms of momentum [19,20], which agrees well with the experimental data [21,22].

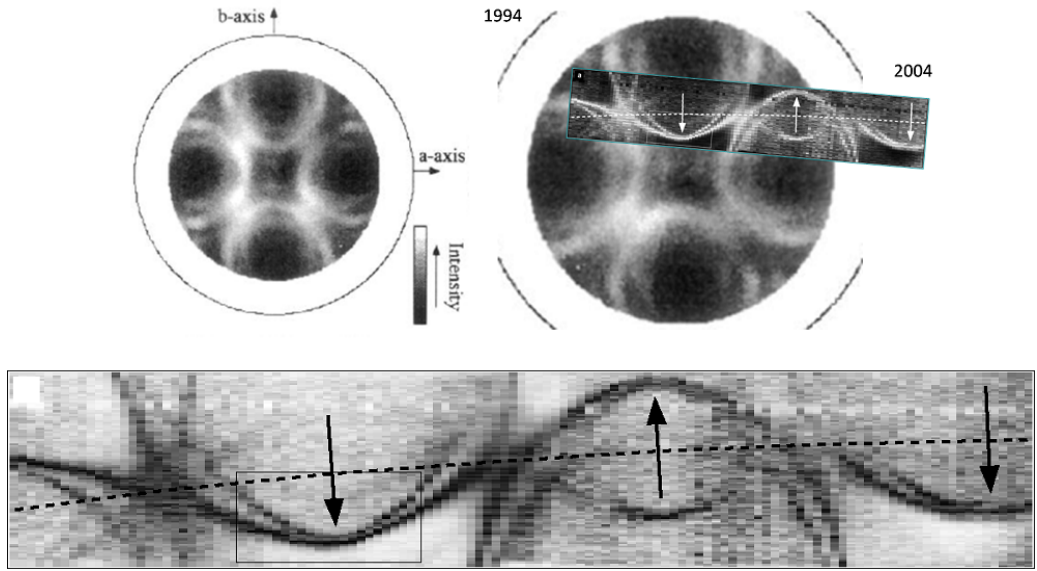

FIG. 3. Examples of the Fermi surface of Bi-2212: from the pioneering work of Aebi [14] measured in 1994 (upper left) using a one-channel analyzer, and measured by us [15] (bottom) 10 years later using a SES 100 analyzer with a multichannel plate detector. Top right: both data sets are shown on the same momentum scale for comparison.

Therefore, in the case of a single or isolated band, the influence of matrix elements can be usually neglected or compensated through a certain renormalization. For example, this has been possible for determining the Fermi surface [23,24], studying the shape of ARPES-spectra [13], determining the self-energy [25], and evaluating relevant interactions [26]. However it is not uncommon, that the variation of the matrix elements with momentum is mistaken for an unusual behavior of $A(\omega, \mathbf{k})$ [27–30]. On the other hand, it is only the dependence of $M$ on the band number that allows to separate the contributions from the neighboring bands [15,21,31] and to unravel complex electronic structures [32–37] by varying the energy and/or polarization of light. The latter requires using the synchrotron radiation with variable energy and polarization [38].

### 2.4. Surface or volume?

A frequently discussed issue is the surface sensitivity of ARPES: From which depth are the electrons emitted? The mean free path of an electron in the crystal as a function of its energy is described by a "universal curve" with a minimum of about 2–5 Å at 50–100 eV [12]. In reality, this dependence is neither universal, nor smooth, i.e., the escape depth is strongly dependent on the material and changes rapidly and non-monotonously upon small variation in energy [39]. Moreover, what is more important is not the escape depth, but how the ARPES spectrum represents the bulk electronic structure of the studied crystal. In most cases, the answers to both questions can be obtained from experiment. Let us consider the examples of specific compounds.

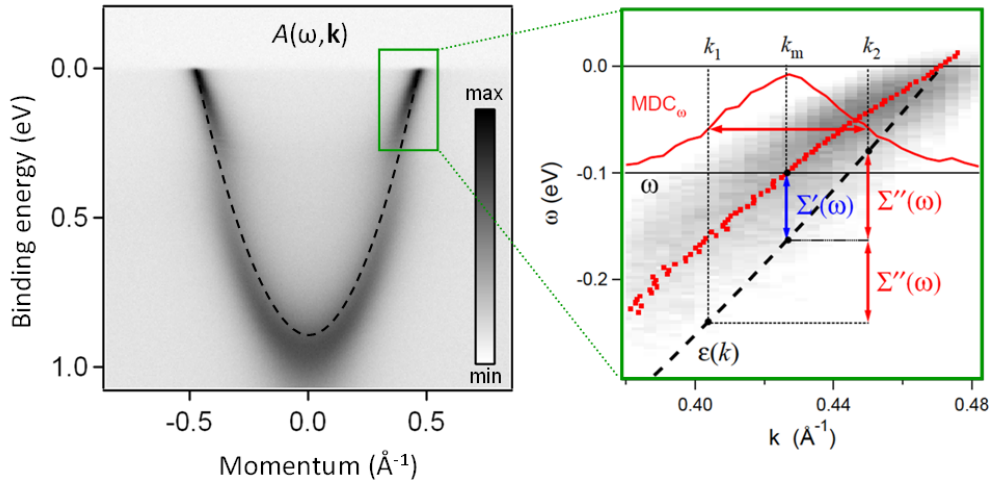

FIG. 4. ARPES-spectrum—an image from the two-dimensional detector of a photoelectron analyzer, which represents, in fact, the one-electron spectral function $A(\omega, \mathbf{k})$. Left: Relation of the experimental distribution of electronic states with the "bare" dispersion of non-interacting electrons, $\varepsilon(k)$, and the self-energy of quasiparticles (one-electron excitations), $\Sigma = \Sigma' + i\Sigma''$.

Among the HTSC cuprates, $Bi_2Sr_2CaCu_2O_{8+\delta}$ (BSCCO or Bi-2212) is the undoubted leader in terms of the sheer number of ARPES studies devoted to it [2,3]. Figs. 2–4 show the spectra for this particular compound. This popularity is due to the presence in its structure of BiO planes bound together by weak van der Waals forces. When a Bi-2212 crystal is cleaved in ultrahigh vacuum, the top plane is BiO, followed by SrO, and only afterwards lies the first conductive double-layer $CuO_2$, in which the superconductivity takes plase [40]. Apparently, it is this double protection that leads to the fact that the observed electronic structure is fully consistent with the bulk one. This can be best proven by the magnitude and temperature dependence of the superconducting gap [2,41].

The depth from which the photoelectrons are emitted in cuprates can be judged from the data on another compound—$YBa_2Cu_3O_{7-d}$ (YBCO) [42]. Excellent surface quality of the cleaved crystals enables high quality ARPES spectra (Fig. 5). However, the superconducting gap is not observed in these spectra. YBCO crystals are cleaved between the BaO and chains-containing $CuO$-planes, thus changing the number of carriers in the latter as well as in the nearest double $CuO_2$ layer, which becomes strongly overdoped and by non-superconducting, as can be judged from the area of the Fermi surface [18,42]. It turned out that the superconducting component, related to the second bilayer of $CuO_2$, can be seen in some spectra [42] and even extracted by using circular polarized light [18]. This suggests that the photoemission in cuprates occurs from the depth of about two unit cells, i.e., ~15 A°.

Despite these difficulties in observation of the bulk electronic structure of YBCO, it can be expected that with the exception of a larger two-layer splitting and chain states [42], the Fermi surfaces of BSCCO and YBCO are very similar. That is why the discovery in 2007 of quantum oscillations [43] with a frequency corresponding to a small Fermi surface was perceived as a contradiction between the "surface" photoemission and "bulk" oscillation methods. However, it soon became clear that the "small Fermi surface" should be an electronic pocket centered around $(0,\pi)$ [44], which may be a simple consequence of either magnetic [45] or crystalline (as in BSCCO) 2x2 superstructure [14,23]. At the moment, considering the number of both new frequencies detected [46] and possible superstructures [47], the situation with the oscillations in YBCO seems to be too complicated to talk about their contradiction with the ARPES data for BSCCO.

Another example of a compound in which both the surface and bulk are visible in the spectra is $Sr_2RuO_4$. Fig. 6 shows a map of the Fermi surface [48], where the splitting of certain sheets can be clearly seen (see the inset). It is interesting that for this compound the observed Fermi surface is in excellent agreement with the band calculations [49] and the measurements of the de Haas–van Alphen oscillations [50].

Three-dimensionality ($k_z$-dispersion) in ferropnictides is much greater than in cuprates. This can be concluded from the observed dependence of the electronic band structure (the size of the Fermi surface) on the photon energy [33,51]. The three-dimensionality highly complicates the lineshape analysis of the spectra, however the fact that $k_z$-dispersion can be observed and even estimated from an ARPES experiment suggests that the escape depth here is significantly larger than two elementary cells. Thus, for the majority of compounds (including the most studied 122 and 111 families [34]) as well as for the BSCCO, the gaps determined using ARPES are in excellent agreement with bulk methods [52].

Thus, there is no clear answer to the question "surface or volume." There are compounds where ARPES "sees the volume" and those where this is hindered by the surface layer ("polar surface") and extremely shallow emission depth. However, most often we can confidently say to which type this particular compound belongs, especially when it comes to superconductors with a gap in the one-electron spectrum.

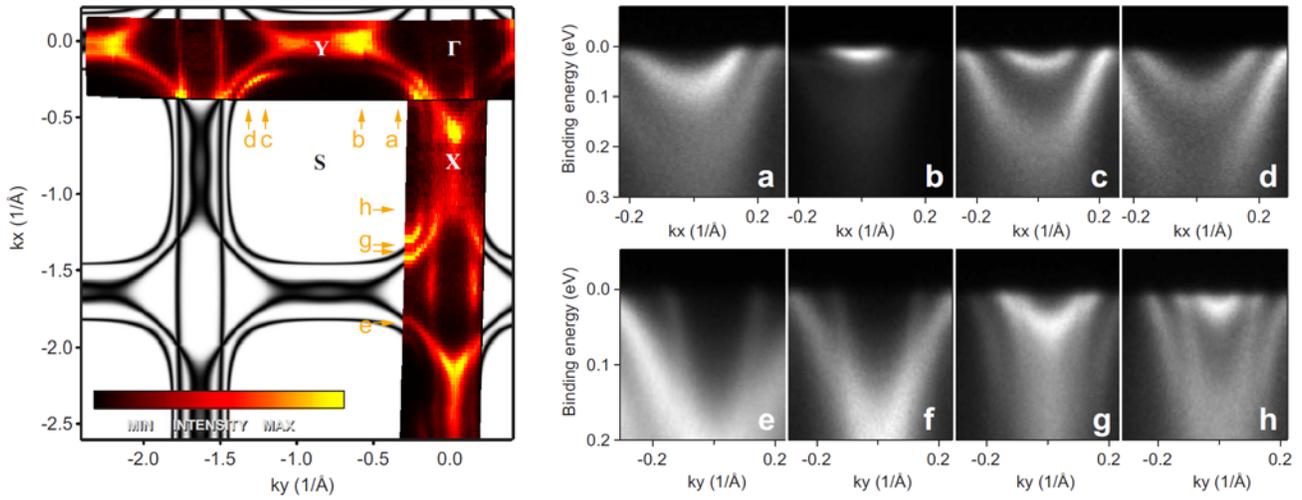

FIG. 5. Experimentally measured electronic structure of untwinned YBCO ($T_c$ = 90 K) [42]. The Fermi surface (left) is represented by two scans (maps) along the perpendicular crystallographic directions laid over the tight-binding model. Arrows indicate the positions of cross-sections (a-h) of the underlying electronic structure; ARPES spectra are shown on the right; $h\nu$ = 50 (a-d) и 55 (e-h) eV, $T$ = 18 K.

To conclude this section, one can say that ARPES spectrum is, in fact, the one-electron spectral function modulated by the matrix elements. The main proof for this statement is the extensive experience: in comparing the ARPES spectra with the electronic band structure calculations; in self-consistent determination of the self-energy [25]; or in comparison with the properties obtained by other methods—the properties that can be also derived from the one-electron spectrum [52,53]. In the following, we consider some examples of this experience.

## 3. ARPES on cuprates—a story of "insight"

In 1987, shortly after the discovery of HTCS, P.W. Anderson published in *Science* a landmark paper [54] in which he has defined that the main features of the new superconductors are their quasi-two-dimensional character and the fact that their superconductivity is formed by doping a Mott insulator [55]. He has predicted that the combination of these features should lead to fundamentally new physics that goes beyond the existing theory of metals [56]. This prediction has been enthusiastically accepted by a number of researchers, and for a long time it was considered indecent to mention such concepts of the solid-state theory as one-particle electronic structure [1] or the Fermi liquid [7] in reference to HTSC [57].

### 3.1. Fermi surface

However, already the first ARPES-experiments on YBCO [58], BSCCO [59], and Nd$_{2-x}$Ce$_x$CuO$_4$ [60] revealed the dispersion and the Fermi surface very similar to those obtained by single-particle calculations (Fig. 7). Although the experimental resolution at that time and other problems [24] did not allow to resolve the Fermi surface splitting [61] or even to argue about its topology [59], these data allowed to speak about the electronic structure of cuprates as a renormalized (twice) one-electron conduction band

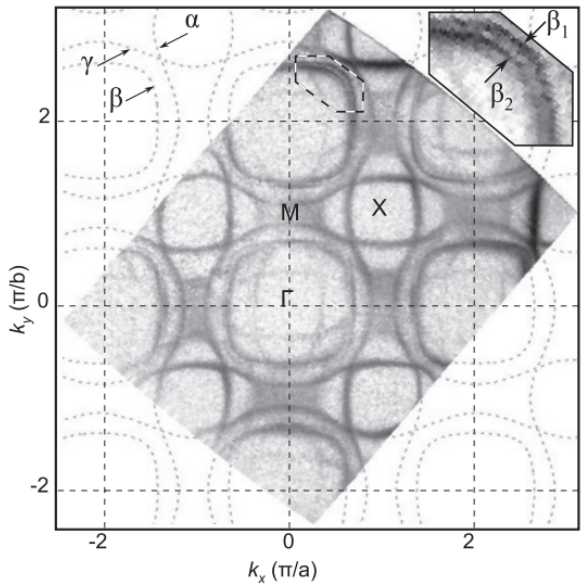

FIG. 6. Experimentally measured Fermi surface of Sr$_2$RuO$_4$, which consists of surface and bulk components [48].

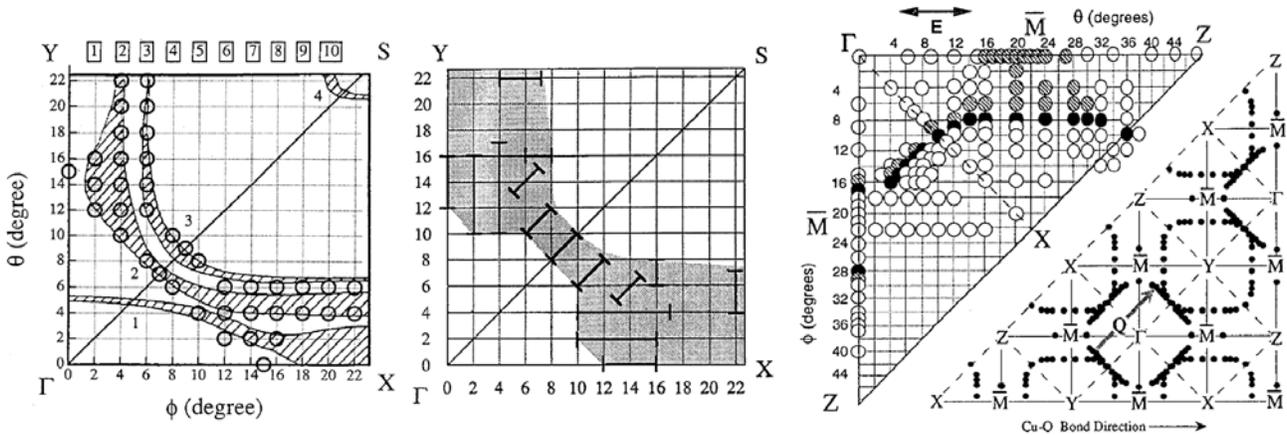

FIG. 7. Fermi surfaces defined in early (early 90s) ARPES-experiments on YBCO [58] (left) and BSCCO [59] (right) using the spectra measured at the indicated points of the Brillouin zone. In the middle: an alternative [61] interpretation of the data of Ref. 58.

formed predominantly by Cu $3d(x^2-y^2)$ and O $2p(x,y)$ orbitals [62].

Further studies, thanks to the rapid development of the method (switching to 2D detectors and improving the energy and angular resolution) and progress in the quality of single crystal growing, have significantly improved visualization of the Fermi surface leaving little room for the "fundamentally new physics." In particular, starting from a certain time [63,64], most of ARPES-groups began to observe the splitting of the conduction band in the bilayer cuprates into the sheets corresponding to the bonding and anti-bonding orbitals, which contradicts the idea of spatial confinement of electrons in separate layers [54,65]. Later, the bilayer splitting was found even along the nodal direction in BSCCO (see Fig. 3) [15], where it is very small.

It has been also shown [24] that the Fermi surface satisfies the Luttinger theorem, i.e., its volume corresponds to the number of conduction electrons per unit cell and is proportional to $(1-x)$, where $x$ is the hole concentration. Moreover, the relative hopping integrals [66] have been determined from the Fermi surface geometry, which defines the geometry of the conduction band and suggests that the onset of the superconducting region in the phase diagram in the direction of reducing the hole concentration starts immediately after the Lifshitz topological transition of the Fermi surface described by the anti-bonding wave function [3].

Determining the Fermi surface in a broad range of momenta allowed to understand the nature of the shadow band [14,55,67,68]. It turned out that this band has a structural origin [67] and is a consequence of the orthorhombic distortions of the tetragonal symmetry of BiO-planes, in the bulk and on the surface of BSCCO [68]. Furthermore, it was shown that another "5x1" modulation of the BiO layer [69] might easily lead to wrong conclusions [70]. For example, in the analysis of the temperature evolution of ARPES-spectra without detailed mapping of the Fermi surface, it has been concluded about the existence of circular dichroism [71]. On the other hand, in Pb-doped BSCCO, in which this modulation is highly suppressed, this effect has not been observed [22,72].

### 3.2. Spectral function

If there are several neighboring bands, as was noted in Sec. 2.3, than the easiest and most effective tool to analyze the structure of ARPES-spectra or to determine the components of the spectral function is the variation of matrix elements by changing the energy and polarization of the light [3,73]. This approach can be well illustrated by the story of "peak-dip-hump" [22]—the line shape of the energy distribution curve (EDC) from the antinodal region around $(0, \pi)$.

Thanks to the influence of Ref. 54 and early ARPES-experiments [65], it was assumed that the bilayer splitting in cuprates is absent, and the corresponding double-hump EDC structure is related exclusively to the strong interaction of electrons with some "mode." However, the observed strong dependence of this structure on the photon energy [21] has indicated that the main reason for that shape is specifically a bilayer splitting. Soon after, those two bands have been observed in ARPES-spectra directly [74]. It has been found that there is indeed an interaction with a "mode," but it is much weaker [74], and its strength depends on the doping level, increasing with decreasing the concentration of holes and disappearing upon overdoping [75]. A so-called magnetic resonance [76], which is a divergence in the spectrum of spin fluctuations [77], has been initially considered as the "mode" in question. However, the same effect was expected from the phonon optical modes [78]. In this respect, an important consequence of elucidating the role of the bilayer splitting was a strong dependence of the interaction with the mode on the doping level, which allowed to argue in favor of the spin-fluctuation mechanism in terms of proximity to the antiferromagnetic undoped compound [3,79]. However, the application of the concepts of quasiparticles, spectral function, and self-energy to cuprates has been often contested and therefore requires experimental validation.

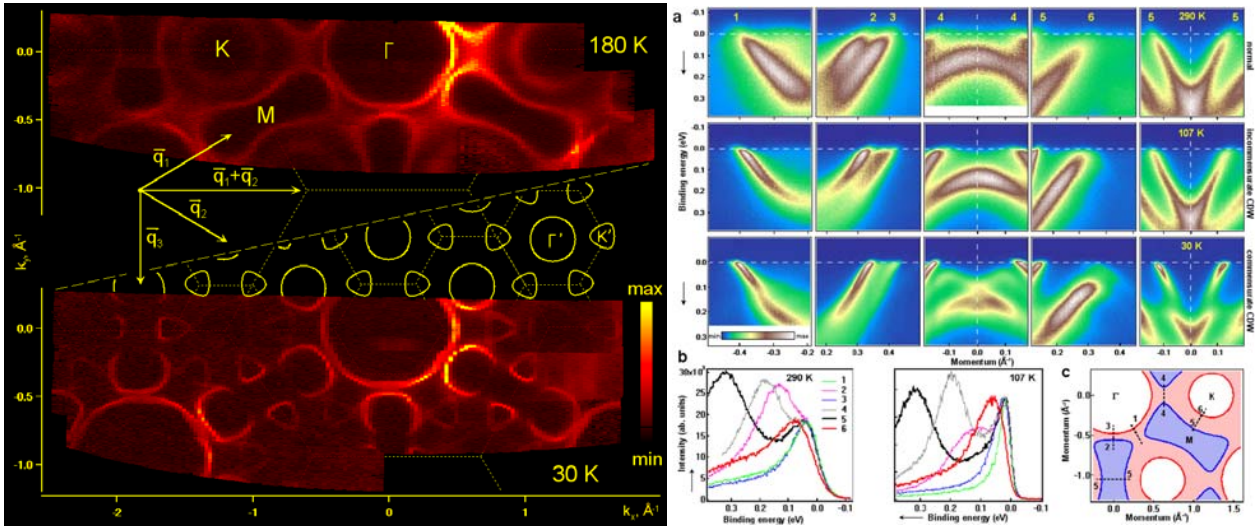

FIG. 8. (left) Fermi surface of $2H$-TaSe$_2$. (right) Details of the electronic structure in the normal state (top row), incommensurate (middle row), and commensurate (bottom row) states (a); single EDC (b); Schematics of the Fermi surface with the respective sections shown (c) [90].

The nodal direction in BSCCO is ideal case to study the applicability of the Green's function formalism to HTSC cuprates. The reasons are: the absence of both the superconducting gap and pseudogap, that gives possibility to restrict ourselves to the normal component of the spectral function; and a simple (similar to parabolic) one-electron dispersion, which is almost degenerate in the case of multi-layer splitting [15]. Moreover, there is a moderate renormalization ($1+\lambda \approx 2$) [80], leading to the "70 meV kink" [81–84], the origin of which has also caused heated debates, most of which can also be attributed to the "phonons or spin fluctuations" dilemma.

From Fig. 4 [25], which illustrates the structure of the spectral function Eq. (1), one can see that the real and imaginary parts of the self-energy at a given frequency $\omega$ are related to the parameters (FWHM) of the momentum distribution curve (MDC) [81] through one-electron dispersion. Thus, in order to determine $\Sigma'(\omega)$ and $\Sigma''(\omega)$ independently from the MDC analysis, one needs to know $\varepsilon(k)$. Since $\Sigma(\omega)$ is an analytical function, all three functions, $\Sigma'(\omega)$, $\Sigma''(\omega)$ and $\varepsilon(k)$ can be determined from the experiment. The algorithm involves finding the parameters $\varepsilon(k)$ such that $\Sigma'(\omega)$ and $\Sigma''(\omega)$ can be expressed in terms of each other using Kramers-Kronig transform [25]. It turned out [25,80] that this procedure does not always work, but only when the ARPES spectrum is originated from one piece of the sample and is clean from experimental artifacts, i.e. really is well described by the spectral function (1). This can be regarded as empirical evidence of applicability of the concept of quasiparticles and Green's function method to superconducting cuprates [79].

As a result of such self-consistent analysis, it has been found [25,80] that for the nodal direction the strength of the key interaction also correlates with proximity to the antiferromagnet. This suggests that the spin-fluctuations are the main contributors to the quasiparticle self-energy throughout the Brillouin zone [79]. A direct comparison of a single-particle ARPES-spectrum with a two-particle spectrum of spin-fluctuations obtained by inelastic neutron scattering for the same YBCO crystals has completely confirmed this hypothesis [85].

So, one may conclude that the HTSC cuprates with the carrier density in the superconducting region of the phase diagram are quasi-two dimensional metals, which are well described by one-electron band structure within the quasiparticle approach, but taking into account a strong electron-electron interaction, the main mediator of which is apparently spin-fluctuations. From the ARPES point of view, the HTSC cuprates were a system that helped to reveal the potential of the method to the fullest extent and contributed to its extremely rapid development.

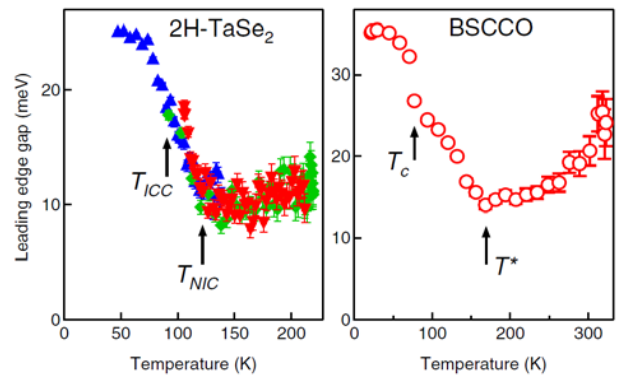

FIG. 9. Temperature dependences of the commensurate band gap/incommensurate pseudogap in $2H$-TaSe2 (left) and the superconducting gap/pseudogap in BSCCO (right) [94].

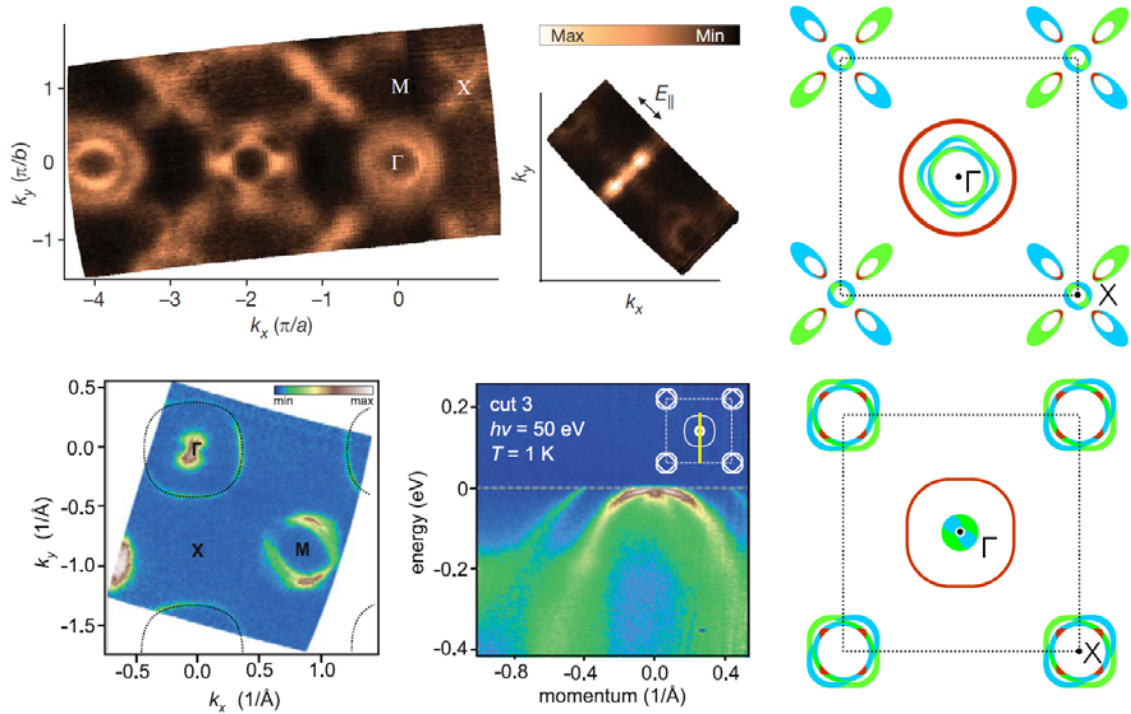

FIG. 10. Experimentally obtained Fermi surfaces of $Ba_{0.6}K_{0.4}Fe_2As_2$ (BKFA) [95] (top row) and LiFeAs [97] (bottom row). Right: Fermi surfaces constructed from the experimental data [35] and labeled in accordance with the main orbital character (see Fig. 11).

## 4. Pseudogap and electronic ordering

In the previous section we did not to raise the issue of the pseudogap intentionally. With respect to cuprates, this phenomenon has been considered in numerous reviews [86,87], but despite the existence of a quite reasonable two-gap scenario [88,89], it has not lost its aura of mystery even today [86,87]. Therefore, in this section, we want to consider the manifestations of the pseudogap phenomenon in other compounds, the transition metal dichalcogenides, briefly mentioning the possible analogy with the cuprates.

Fig. 8 shows the Fermi surface of $2H$-TaSe$_2$ [90], a compound in which there are two phase transitions into the states with incommensurate (122 K) and commensurate 3x3 (90 K) charge-density wave (CDW). Moreover, it is the first transition at which a jump in the heat capacity and a kink in the resistance are observed, while the second transition has almost no effect on these properties [91]. From ARPES point of view the situation is opposite. The Fermi surface, shown above, remains virtually unchanged up to 90 K, and a new order appears just below the commensurate transition.

The explanation for this "paradox" is the behavior of the spectral weight near the Fermi level on the Fermi surface sheet centered around K-points (Fig. 8(c)). Below 122 K the spectral weight starts to decrease sharply, that is the pseudogap opening (see the cross-section 5-6 in Figs. 8(a) and 9). When passing through 90 K, the pseudogap is transformed into a band gap in the new Brillouin zone, but this transition is not accompanied by such a gain in kinetic energy.

These data: (1) prove empirically that the formation of the incommensurate charge density wave leads to a redistribution of the spectral weight at the Fermi level and below, while the transition from incommensurate to commensurate order leads rather to a redistribution of the spectral weight in momentum; and (2) show that the photoemission intensity depends not only on the photoemission matrix elements but also on the type and magnitude of the new order parameter [92].

It is interesting to note that the incommensurate gap in dichalcogenides [90,93] is completely analogous to the pseudogap in cuprates, both in terms of their spectroscopic manifestations (magnitude and anisotropy) and temperature dependence [94] (Fig. 9). Therefore, a similar scenario can be assumed for cuprates, when a pseudogap in the antinodal region appears due to the formation of the incommensurate spin order (the same argument of proximity to the antiferromagnet), which, in turn, is determined by nesting of the straight sections of the Fermi surface [94].

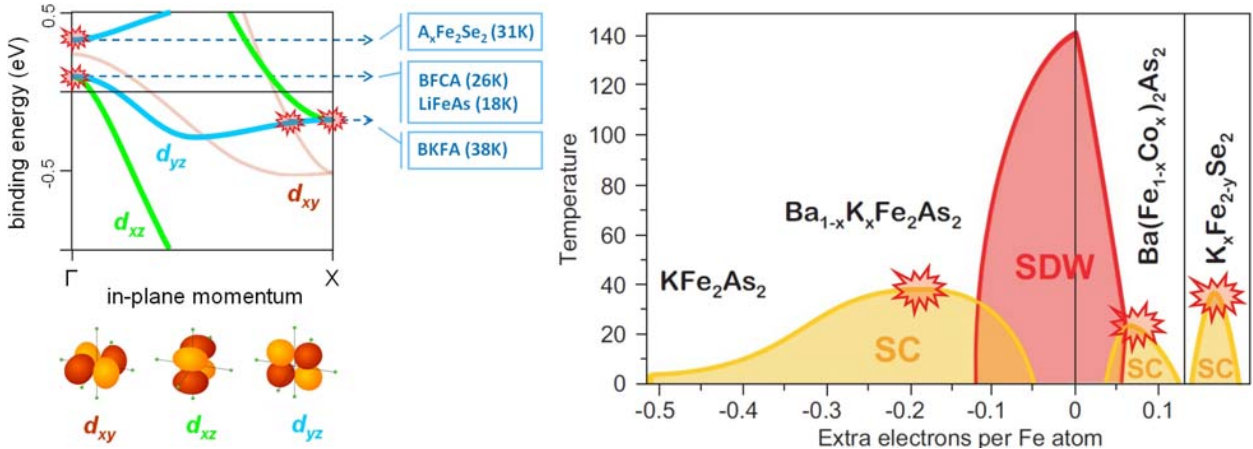

FIG. 11. Generalized electronic band structure (left) and phase diagram of the iron-based superconductors (right). Maximum $T_c$ is observed in the vicinity of the Lifshitz transition, when the top or bottom of one of the bands with $dxz$ or $dyz$ orbital character becomes close to the Fermi level [35].

## 5. Topological Lifshitz transition

Iron-based superconductors (Fe-SC) is a new numerous class of superconductors, which offer interesting physics and look promising in terms of possible applications. It is expected, mainly due to their diversity and similarity to the cuprates, that these compounds will help somehow to solve the mystery of high-temperature superconductivity. Nevertheless, while the pairing mechanism and even the symmetry of the order parameter remain the subject of active debate [34], it is certainly the complexity of the electronic structure (five conduction bands instead of a single band in cuprates) that can give a key to understanding the mechanism of superconductivity in this class of compounds and to increasing the transition temperature [35].

Although the electronic structure of Fe-SC is complex, it is common to all of them. The differences consist of small (of the order of 0.1 eV, but critical for the geometry of the Fermi surface [35]) relative shifts of individual bands and changes in the chemical potential with doping. Thus, there is a unique opportunity to establish correlations between the characteristics of the electronic structure and various properties (primarily, the transition temperature). For example, Fig. 10 on the left shows the experimental Fermi surface of $Ba_{0.6}K_{0.4}Fe_2As_2$ (BKFA) [95, 96] (top row) and LiFeAs (Ref. 96) (bottom row) and, on the right, the Fermi surface constructed from the experimental data and labelled according to their main orbital character (see Fig. 11). It turned out that for all the compounds investigated by ARPES, the maximum $T_c$ is observed in the vicinity to the Lifshitz transition, when the top or bottom of one of the bands with the $d_{xz}$ or $d_{yz}$ symmetry becomes close to the Fermi level [34, 35] (see Fig. 11).

The observed correlation indicates that in order to explain the mechanism of pairing in Fe-SC, the standard BCS model is clearly not enough: the superconducting transition temperature correlates mainly with the geometry of the Fermi surface [34,35], rather than with the density of states at the Fermi level. On the other hand, the density of states is certainly important for superconductivity, so, the obtained correlation indicates a way to increase $T_c$: hole overdoping of $KFe_2As_2$ or LiFeAs.

## 6. Conclusions

Modern ARPES experiment allows to observe directly the electronic structure (the structure of single-particle excitations) in quasi-2D crystals. This shifts the fermiology to the domain of everyday experience and contributes to the empirical (in a positive sense) understanding of the mechanisms that determine the electronic properties of solids. In this paper, it is illustrated by several examples, including HTSC cuprates, where the study of the structure of ARPES spectra and their relation to the spectrum of spin fluctuations have helped to distinguish the latter as the main mechanism of scattering and superconducting pairing. The example of transition metals dichalcogenides shows most clearly the relation between the Fermi surface geometry and the manifestation of instability of the electronic system with respect to the formation of charge density waves. In the case of iron-based superconductors, the complexity of their electronic structure has allowed us to establish an empirical correlation between this structure and superconductivity.

*Translated from: Fiz. Nizk. Temp.* **40**, *375-388 (2014).*